\documentclass[%
 reprint,
 superscriptaddress,
 amsmath,amssymb,
 aps,
]{revtex4-2}

\usepackage{graphicx}
\usepackage{subcaption}
\usepackage{amsmath}
\usepackage{siunitx}

\DeclareSIUnit\arcsec{arc sec}
\DeclareSIUnit\Joule{J}
\DeclareSIUnit\sterradian{sr}
\usepackage{xcolor}
\usepackage{dcolumn}
\usepackage{bm}
\usepackage{tabularx}

\begin{document}


\title{Nanoscale X-ray imaging with high spectral sensitivity using fluorescence intensity correlations}

\author{Tamme Wollweber}
\affiliation{Max Planck Institute for the Structure and Dynamics of Matter, 22761 Hamburg, Germany}
\affiliation{Center for Free-Electron Laser Science, 22761 Hamburg, Germany}
\affiliation{The Hamburg Center for Ultrafast Imaging, 22761 Hamburg, Germany}

\author{Kartik Ayyer}
\email{kartik.ayyer@mpsd.mpg.de}
 
\affiliation{Max Planck Institute for the Structure and Dynamics of Matter, 22761 Hamburg, Germany}
\affiliation{Center for Free-Electron Laser Science, 22761 Hamburg, Germany}
\affiliation{The Hamburg Center for Ultrafast Imaging, 22761 Hamburg, Germany}


\begin{abstract}
This paper introduces Spectral Incoherent Diffractive Imaging (SIDI) as a novel method for achieving dark-field imaging of nanostructures with heterogeneous oxidation states. With SIDI, shifts in photoemission profiles can be spatially resolved, enabling the independent imaging of the underlying emitter distributions contributing to each spectral line. In the X-ray domain, this approach offers unique insights beyond the conventional combination of diffraction and X-ray Emission Spectroscopy (XES). When applied at X-ray Free-Electron Lasers (XFELs), SIDI promises to be a versatile tool for investigating a broad range of systems, offering unprecedented opportunities for detailed characterization of heterogeneous nanostructures for catalysis and energy storage, including of their ultrafast dynamics. 

\end{abstract}

\maketitle


Coherent diffractive imaging (CDI) is a family of techniques which use the interference pattern produced by the elastic scattering of a coherent light field by the sample to determine its structure~\cite{chapman2010coherent,miao2015beyond,pfeiffer2018x}. These lensless techniques are especially powerful for X-ray imaging since one can obtain much better resolution images than limited by the numerical aperture of the optical system. In these measurements, incoherent scattering processes, like fluorescence, have conventionally been considered detrimental due to the absence of a static interference pattern in the far field~\cite{slowik2014incoherent}. Recent advancements, notably the emergence of X-ray free electron laser (XFEL) sources featuring ultra-short pulse durations, have given rise to a novel imaging technique, termed Incoherent Diffractive Imaging (IDI)~\cite{classen2017incoherent}. In contrast to CDI, which presupposes a fixed phase relation between incoming and scattered photons, IDI enables high-resolution imaging through the correlation of only transiently coherent fluorescence photons.

Given the random initial phases of each emitted photon, fluorescence typically results in a uniform intensity distribution in the far field. Thus, the integrated far field does not contain any structural information. However, when fluorescence is detected within its coherence time $\tau_{c}$, i.e. the time interval during which the relative phases are stable, stationary speckle patterns can be detected by means of second-order spatial intensity correlations~\cite{classen2017incoherent}. In pioneering proof-of-principle experiments, copper K$_\alpha$ fluorescence has been measured to determine the focus profile and the pulse duration of an XFEL pulse~\cite{nakamura2020focus, inoue2019determination}, emulating the original astronomical experiments of \textcite{brown1956test}. \textcite{trost2023imaging} recently demonstrated the experimental feasibility of imaging non-trivial 2D structures with this approach. In the optical domain, this method has also been demonstrated to image trapped ion structures~\cite{richter2021imaging}.

IDI has potentially a number of benefits compared to other methods in the imaging of nanostructures. The use of core-level fluorescence results in extremely high element sensitivity, enabling dark-field imaging of the substructure of just a single element in a heterostructure. For the same experimental geometry, IDI can access twice the spatial frequencies as elastic scattering, and in comparison to other element specific techniques, IDI fills a critical resolution gap between local probes like extended X-ray absoprtion fine structure (EXAFS) and lower resolution X-ray fluorescence microscopy (XFM).


However, the signal-to-noise ratio (SNR) demands in IDI are more exacting and was comprehensively studied by \textcite{trost2020photon}. It was found to depend on the number of detected photons per frame, the number of modes present in the emitted light field, and the size and shape of the emitting object. Additionally, the detector position imposes limitations on the sample size transverse to the sample-detector direction due to the finite coherence length associated with the fluorescent emission~\cite{lohse2021incoherent}. If the transverse extent of the sample exceeds the coherence length $L_T = c \cdot \tau_c$, the contrast in IDI will be reduced significantly. Thus, the requirements imposed on the sample structure and the amount of data required result in significant hurdles to its practical application to a broad range of systems.


\begin{figure}
  \centering
  \includegraphics[width=0.95\columnwidth]{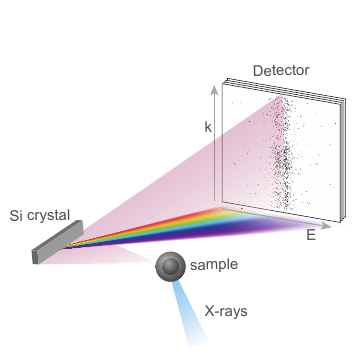}
  \caption{Schematic of the Spectral Incoherent Diffractive Imaging (SIDI) setup. The silicon crystal analyzer disperses the emitted photons in one direction while acting as a simple mirror along the orthogonal dimension. Thus, the position of a detected photon on the two-dimensional detector in the far field corresponds to its energy and one component of its wave-vector.}
  \label{fig:setup}
\end{figure}

In this letter, we propose a method to overcome these experimental challenges and, furthermore, enable imaging that exhibits sensitivity to spectral line shifts associated with different oxidation states of the underlying emitters. We call this technique Spectral Incoherent Diffractive Imaging (SIDI). The underlying principle involves substituting one dimension of the wave vector $\mathbf{k}$, with the photon energy E. Figure~\ref{fig:setup} illustrates how this can be achieved by introducing an analyzer crystal between the sample and the detector, effectively acting as a mirror along one of the $\mathbf{k}$ dimensions, while employing Bragg reflection along the other dimension to split the photons according to their energy, a mechanism analogous to X-ray emission spectroscopy (XES). 

XES has become an essential tool for investigating electronic transitions across various research fields such as physics, material science and environmental sciences~\cite{zimmermann2020modern}. Particularly, $3d$ transition metals are frequently studied using XES to analyze K$_\beta$ line intensities and shifts corresponding to different oxidation states when bonding to different ligands~\cite{zimmermann2020modern,geoghegan2022combining,martin2016kbeta}. This is in particular interesting for the search of new catalysts or more efficient energy storage materials. While the combination of XES with conventional CDI or crystallography~\cite{kern2013simultaneous,makita2023combining} allows simultaneous imaging of the emitter distribution and the recording of emission spectra, elastic scattering cross-sections are only very weakly sensitive to changes in oxidation states and so only information about the relative populations of different oxidation states is obtained. In conventional IDI, where the entire emission spectrum is combined, this leads to loss of contrast since the correlation of photons with different energies adds to the background. This problem was circumvented to some extent in prior experiments by filtering out the K$_\beta$ emission and correlating just the oxidation-state-insensitive K$_\alpha$ photons~\cite{inoue2019determination,trost2023imaging}.

\begin{figure}[ht]
  \centering
  \begin{tabularx}{0.9\columnwidth}{Xc}
    \parbox[c]{0.5\columnwidth}{\includegraphics[width=0.35\columnwidth]{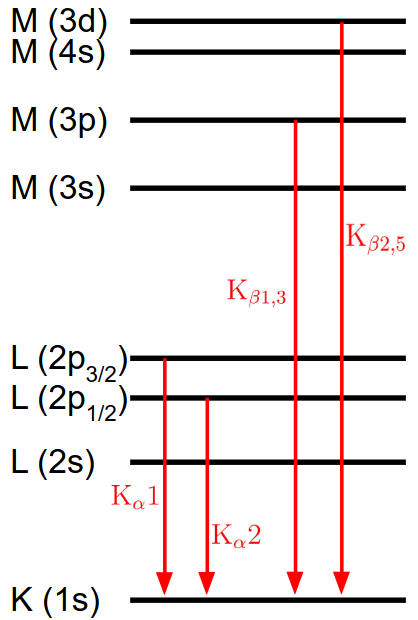}} &
    \begin{tabular}{c|c @{\hspace*{12pt}}c}
      & E & $\Delta$E \\
      \hline
      K$_{\alpha 1}$ & 5898.8 & 2.47 \\
      K$_{\alpha 2}$ & 5887.6 & 2.92 \\
      K$_{\beta 1,3}$ & 6490.0 & 2.97 \\
    \end{tabular} \\
  \end{tabularx}
  \caption{Schematic core electronic structure of manganese. The table on the right specifies the transition energies and line widths of neutral Mn in electron volts~\cite{holzer1997k,peng1994high}.}
  \label{fig:elevels}
\end{figure}

Our proposed approach provides the flexibility to position the detector strategically, enabling the acquisition of specific regions of interest, such as only the K$_\alpha$ or even the more sensitive K$_{\beta 1,3}$ and K$_{\beta 2,5}$ lines (see Fig.~\ref{fig:elevels}). This facilitates the study of structures with heterogeneous oxidation states, where only subtle shifts in the emission spectra occur. Leveraging the potential for time-resolved measurements at XFELs, our method opens the possibility of simultaneously capturing spatially and spectrally resolved movies of redox reactions in complex systems.

In order to illustrate the principle of spectrally resolved imaging of nanostructures, consider first a spherical Mn nanoparticle, which is exposed to a \SI{6.2}{\femto\second} long, \SI{6580}{\eV} self-amplified spontaneous emission (SASE)  XFEL pulse. A flat silicon analyzer crystal and the detector are positioned to satisfy the Bragg condition for the silicon $422$ peak concerning the K$_{\alpha 1}$ and K$_{\alpha 2}$ lines (refer to emission properties in Fig.~\ref{fig:elevels}). We expect \SI{4000} photons per exposure on a $1024\times1024$ pixel detector with a pixel size of \SI{75}{\micro\meter} at an effective detector distance of \SI{1}{\meter}. Details about signal estimation are described in Appendix~\ref{app:signal}.

In conventional IDI, the K$_{\alpha}$ fluorescence of Mn is characterized by a coherence time of $\tau_\text{c} =$ \SI{446}{\atto\second}, given by twice the radiative lifetime $\hbar/\Gamma$~\cite{goodman2015statistical}, where $\Gamma$ represents the spectral linewidth of \SI{2.92}{\eV}~\cite{holzer1997k}. Considering our experimental geometry and small sample volume, the energy resolution is constrained primarily by the pixel size of the detector, which results in an effective resolution of \SI{295}{\milli\eV} and a corresponding increase of the coherence time to \SI{4.4}{\femto\second} and the coherence length to \SI{1.3}{\micro\meter}. The energy resolution can be increased by an order of magnitude with the use of back-scattering analyzers as employed in resonant inelastic X-ray scattering (RIXS) experiments~\cite{gog2013spherical}\footnote{In order to retain one $\mathbf{k}$ dimension, the surface needs to have a cylindrical profile rather than a spherical one.}.

\begin{figure*}
\centering
\begin{tabular}{cc}
  \parbox[c]{0.32\linewidth}{\includegraphics[width=0.99\linewidth]{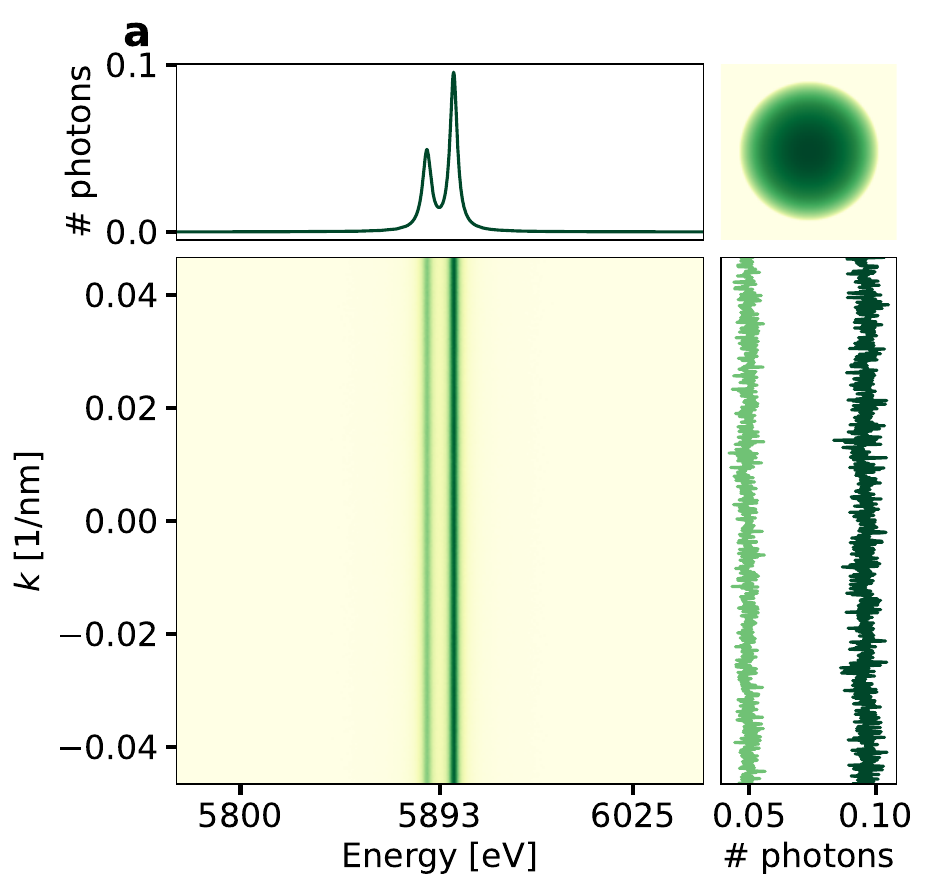}} &
  \begin{tabular}{ccc}
    \includegraphics[width=0.2\linewidth]{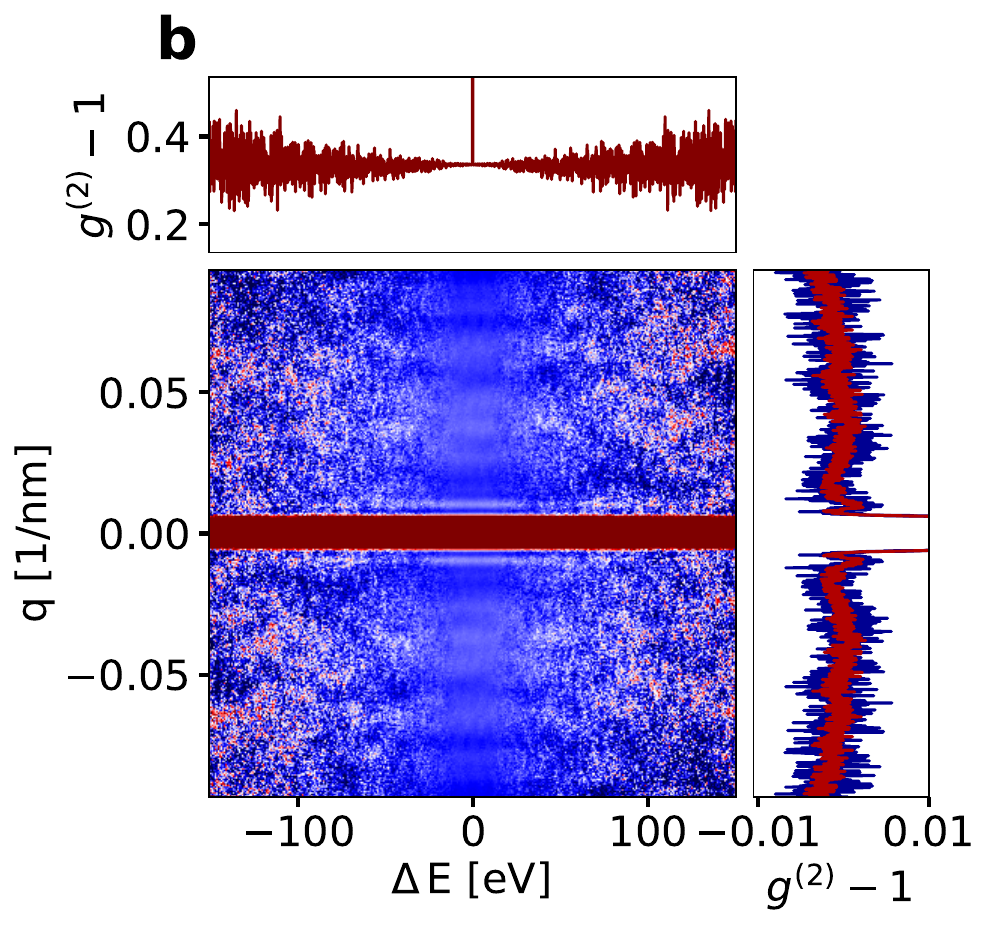} &
    \includegraphics[width=0.2\linewidth]{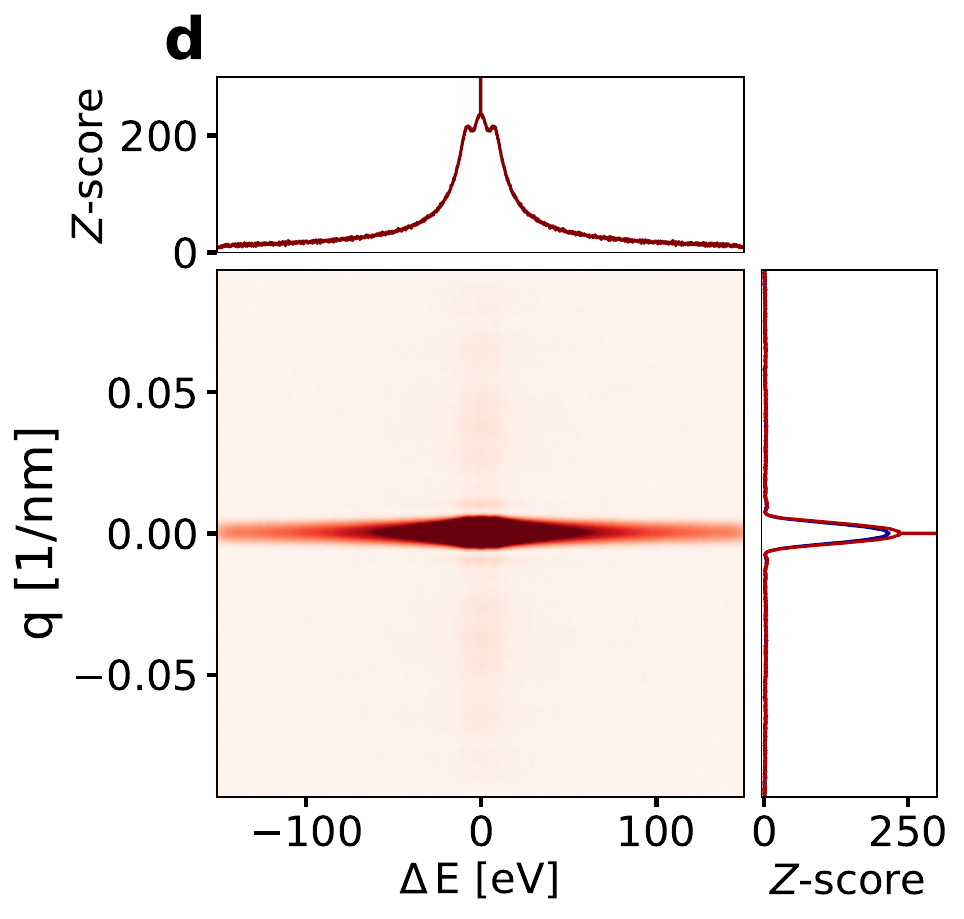} &
    \includegraphics[width=0.2\linewidth]{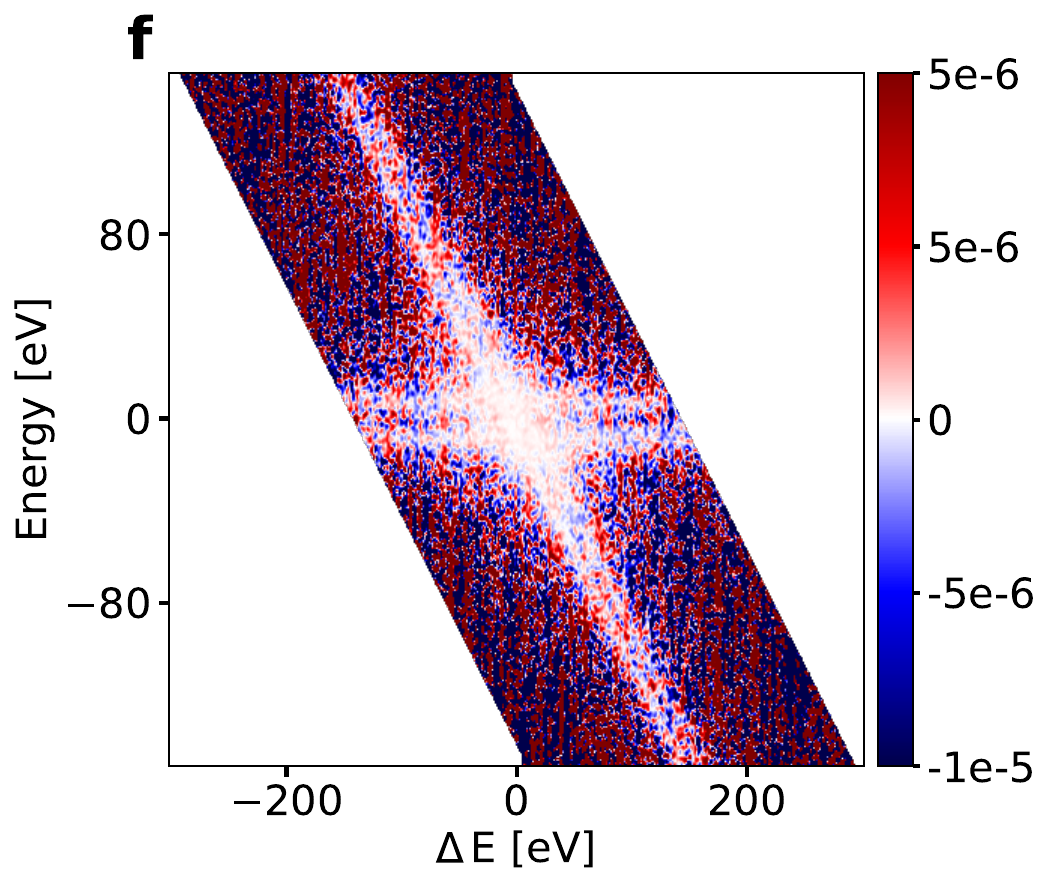} \\
    \includegraphics[width=0.2\linewidth]{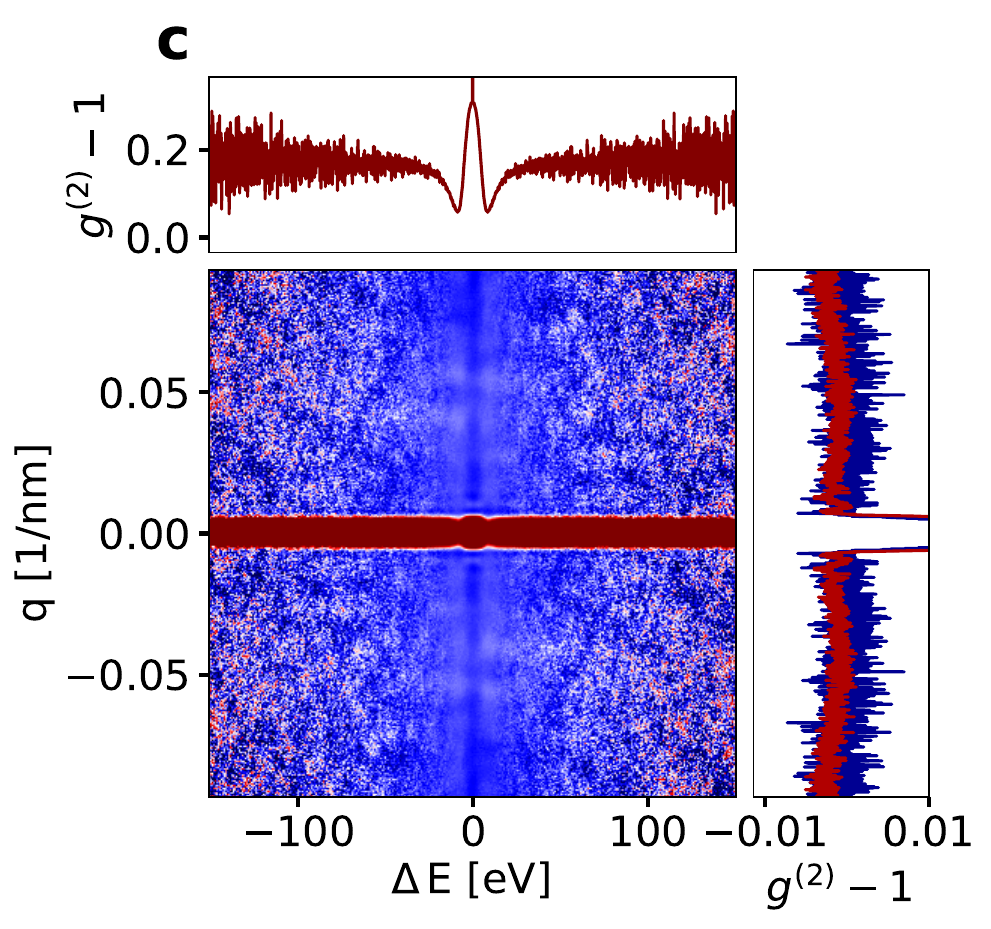} &
    \includegraphics[width=0.2\linewidth]{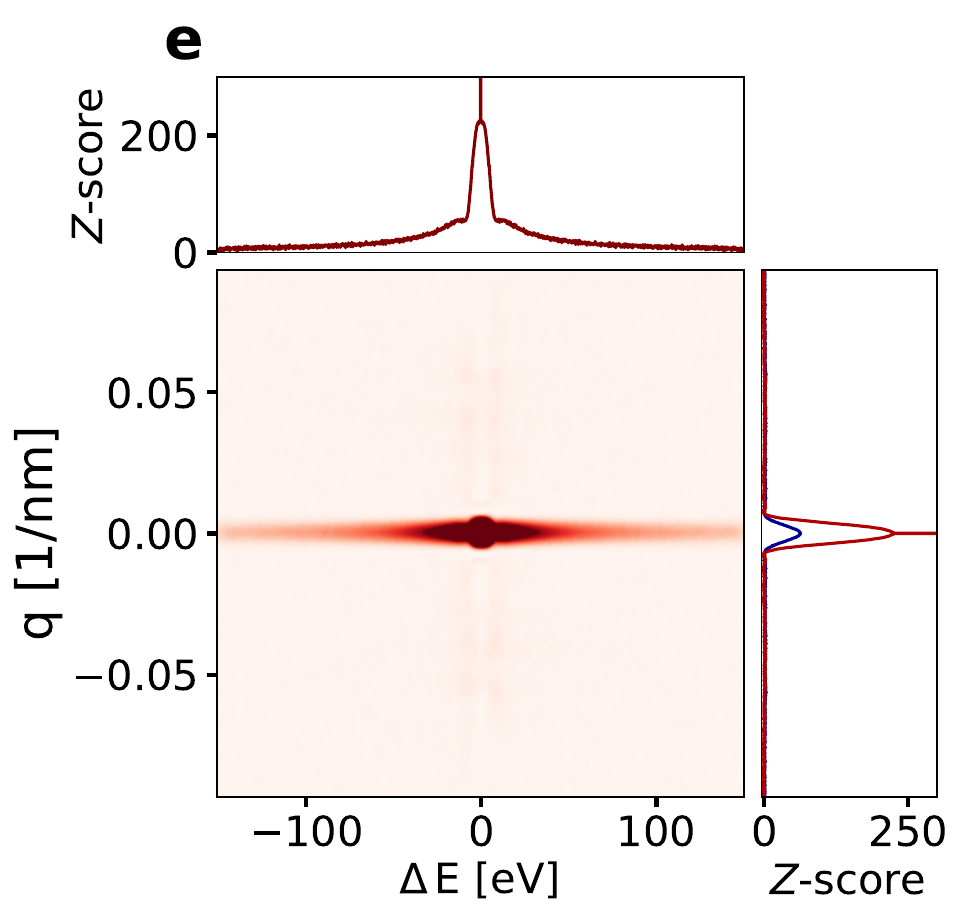} &
    \includegraphics[width=0.2\linewidth]{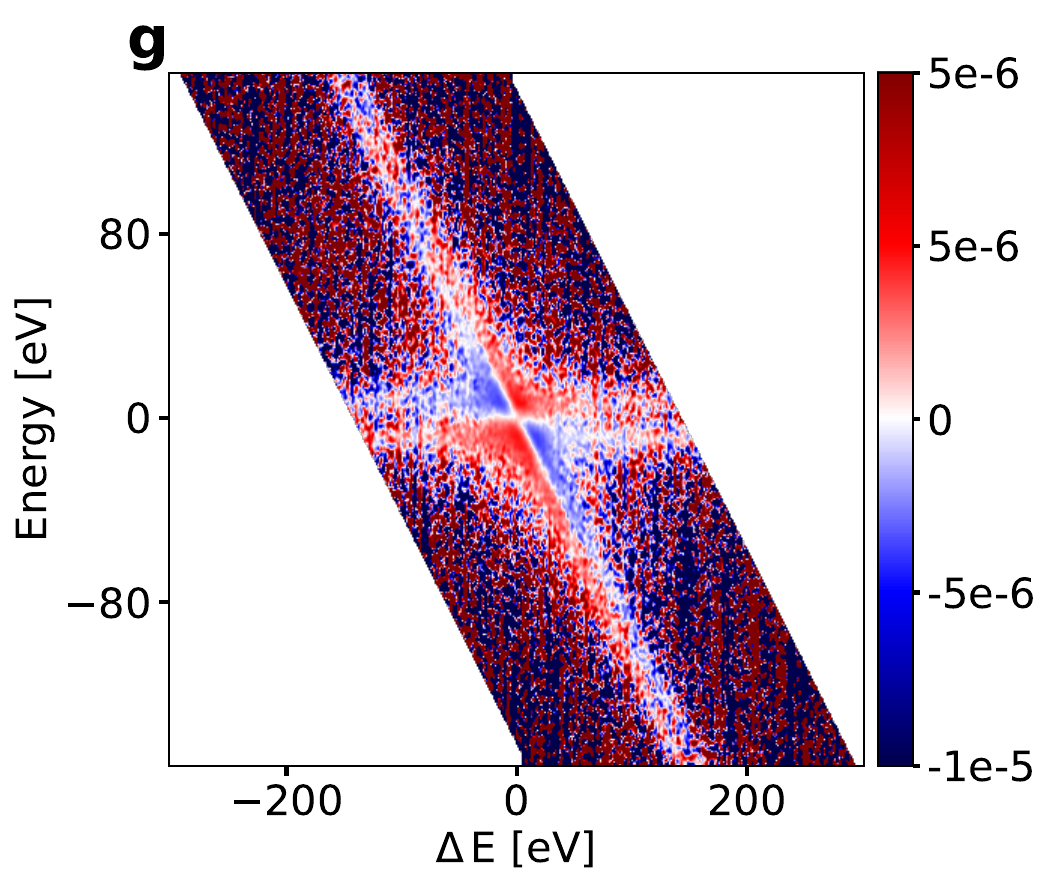} \\
  \end{tabular}
\end{tabular}
\caption{K$_\alpha$ spectral IDI simulations. (\textbf{a}) Average 2D detector image showing the two Mn-K$_\alpha$ lines. As expected, there is no structure along the $k$ axis. The inset in the upper right shows the projection of the spherical nanoparticle. (\textbf{b,~c}) $g^{2}(q, \Delta E)$ images obtained by averaging 2D intensity correlations for each exposure. The top row shows the behavior for full phase correlation between the K$_{\alpha1}$ and K$_{\alpha2}$ wavefunctions, while the bottom row shows the case of no phase correlation. (\textbf{d,~e}) $Z$-score for the images in \textbf{b} and \textbf{c} respectively. The vertical profiles in (\textbf{b \textendash ~e}) correspond to the central line cuts and the vertical profiles to the central (red) and the off-central cuts (blue) at $\Delta E = E_{\text{K}_{\alpha1}}-E_{\text{K}_{\alpha2}}$. (\textbf{f,~g}) $g^{(2)}(E, \Delta E)$ signal for $q =$ \SI{3.28}{\per\micro\meter} showing a clear difference between the two cases.}
\label{fig:kalpha}
\end{figure*}

In order to retrieve the Fourier amplitudes $|F(\mathbf{q})|$ from the flat intensity profile along $\mathbf{q}$, we employ the well-known Siegert relation~\cite{goodman2015statistical}, calculating the second-order cross-correlation:

\begin{equation}
\label{eq:g2}
\begin{split}
    g^{(2)}(\mathbf{q}) &= \frac{\left \langle I (\mathbf{k}) \cdot I(\mathbf{k} + \mathbf{q}) \right \rangle }{\left \langle I(\mathbf{k}) \right \rangle ^2} \quad = \quad 1 + \beta \left | g^{(1)}(\mathbf{q}) \right | ^2 \\
    &= 1 + \beta \frac{\left | F(\mathbf{q}) \right |^2}{\left | F(0) \right | ^ 2}
\end{split}
\end{equation}
Here, $\beta$ is the so-called visibility factor determining the contrast of the correlation function, primarily influenced by the number of temporal modes and their photon occupancy within a single exposure~\cite{trost2020photon}. The statistical significance of the measured $g^{(2)}$ value is then expressed through the $Z$-score:

\begin{equation}
\label{eq:zscore}
        Z{\text{-score}}(\mathbf{q}) = \frac{\left \langle I (\mathbf{k}) \cdot I(\mathbf{k} + \mathbf{q}) \right \rangle \cdot \sqrt{N}}{\left \langle \left ( I (\mathbf{k}) \cdot I(\mathbf{k} + \mathbf{q}) \right ) ^2 \right \rangle - \left \langle I (\mathbf{k}) \cdot I(\mathbf{k} + \mathbf{q}) \right \rangle ^2} \, ,
\end{equation}
where $N$ is the number contributing frames. To enhance contrast, the second-order intensity correlation should be calculated independently for each energy bin, corresponding to each pixel column on the detector. Although one $\mathbf{q}$ component was substituted by the energy dimension, a two-dimensional $g^{(2)}$ calculation can still be performed, yielding correlations between the different energies, $g^{(2)}(q,\Delta E)$, where $q$ refers to the undispersed component of $\mathbf{q}$. In particular, we can understand the relative phase correlation of photons in the different K$_\alpha$ sub-levels.

Figure~\ref{fig:kalpha}a displays the integrated far-field intensity for the two K$_{\alpha_{1,2}}$ lines after \SI{10000} exposures, showing well-separated flat intensity profiles along $q$, as expected. See Appendix~\ref{app:simulation} for details about the simulation. Summing the integrated intensity along $q$ yields the emission spectrum similar to a conventional XES experiment as seen in the horizontal profile in Fig.~\ref{fig:kalpha}a. 

But the intensity correlation approach can provide more insight into the coherence of the quantum evolution of the system following the creation of the $1s$ core hole. If the decay occurs before decoherence due to coupling with the environment, the measured intensity will be a coherent combination of the two photon wavefunctions for each spectral line,
\begin{equation}
    \label{eq:superposition}
    I \propto \left | \Psi \right | ^2 = \left | \psi_{\alpha1} + \psi_{\alpha2} \right | ^ 2,
\end{equation}
This result differs fundamentally from the case where emission from the two different decay channels combines \emph{incoherently},
\begin{equation}
    \label{eq:inc_sum}
    I \propto \left | \psi_{\alpha1} \right | ^2 + \left | \psi_{\alpha2} \right | ^ 2,
\end{equation}

By correlating the photon positions between the two spectral lines, the two-dimensional $g^{(2)}(q,\Delta E)$ can be used as a measure of the phase coherence, and thus the degree of superposition of the two quantum states $L_1 (2_{\text{p},1/2})$ and $L_3 (2_{\text{p},3/2})$ associated with these emission lines (Fig.~\ref{fig:elevels}). The corresponding two-dimensional spatial intensity correlations are shown in Fig.~\ref{fig:kalpha}b and \ref{fig:kalpha}c for the coherent (Eq.~\ref{eq:superposition}) and the incoherent (Eq.~\ref{eq:inc_sum}) sum of the wave functions, respectively.

The shape of the central vertical profile remains the same in both cases, as seen in the sum of the autocorrelation for each detector column. However, the inter-level components in Fig.~\ref{fig:kalpha}b distinctly reveals $g^{(2)}$ correlations for $\Delta E = E_{\text{K}_{\alpha1}}-E_{\text{K}_{\alpha2}}$, which are absent when the wave functions are added incoherently. In both cases, there are weak first order fringes visible along the $q$ dimension. However, the $Z$-score is only significant for the central speckle, as shown in Fig.~\ref{fig:kalpha}d and Fig.~\ref{fig:kalpha}e. In order to achieve higher $g^{(2)}$-contrast and better SNR, it would be necessary to increase the number of photon counts or to acquire more frames~\cite{trost2020photon}.

As already shown in Figs.~\ref{fig:kalpha}b and \ref{fig:kalpha}c, SIDI yields the second degree of coherence for different energies. In fact, the underlying data can be analyzed as a three-dimensional correlation function $g^{(2)}(q,E,\Delta E)$, which is now also a function of the relative difference between two energies. Figures~\ref{fig:kalpha}f and \ref{fig:kalpha}g show the correlation $g^{(2)}(E,\Delta E)$ for a fixed $q$. As expected, there is no energy dependency for the coherent case where one has full phase correlation between both lines, while there is a clear decrease in second-order correlations for larger $\Delta E$ in the incoherent case.

\begin{figure}
\centering
\begin{tabular}{cc}
  \parbox[c]{0.475\linewidth}{\includegraphics[width=0.99\linewidth]{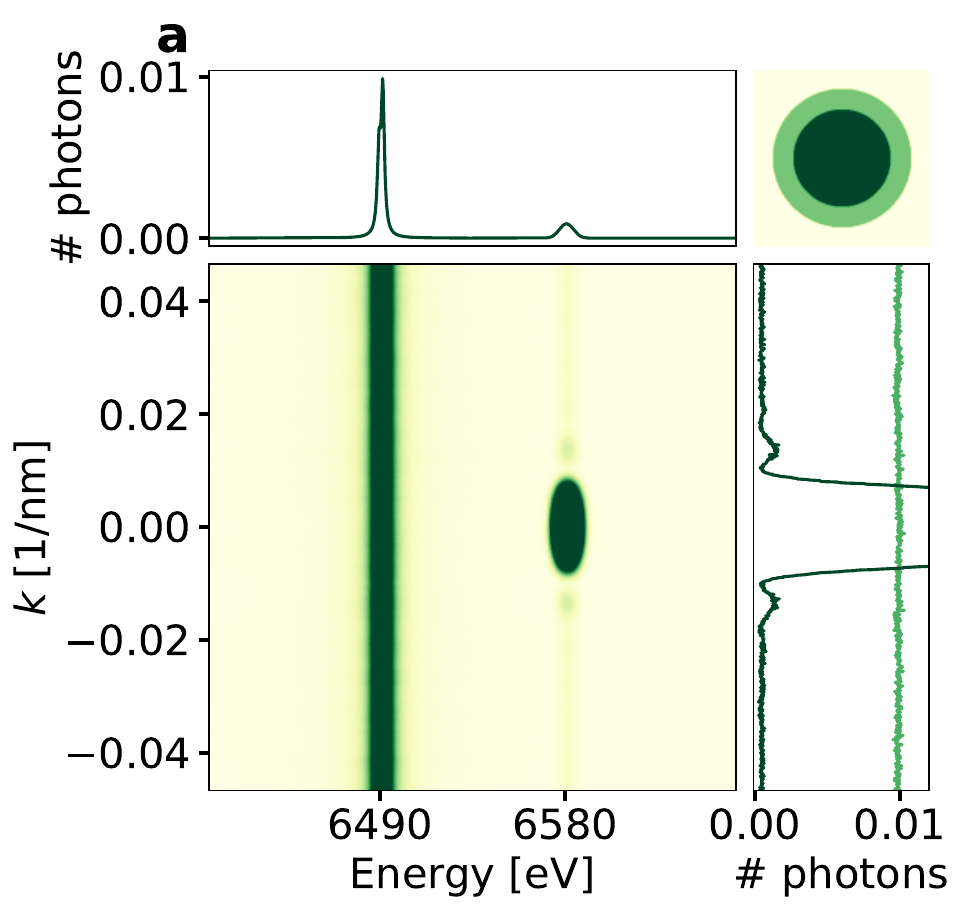}} &
  \begin{tabular}{c}
    \includegraphics[width=0.475\linewidth]{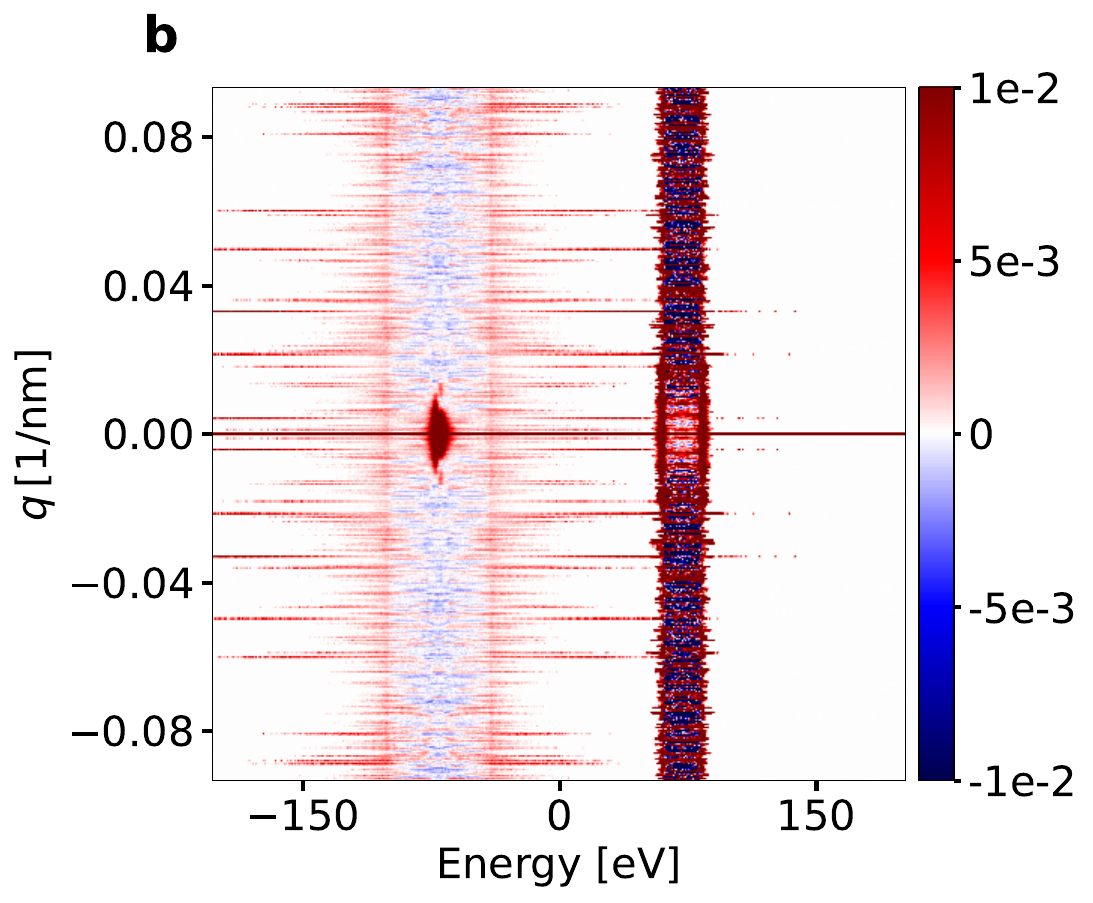} \\
    \includegraphics[width=0.475\linewidth]{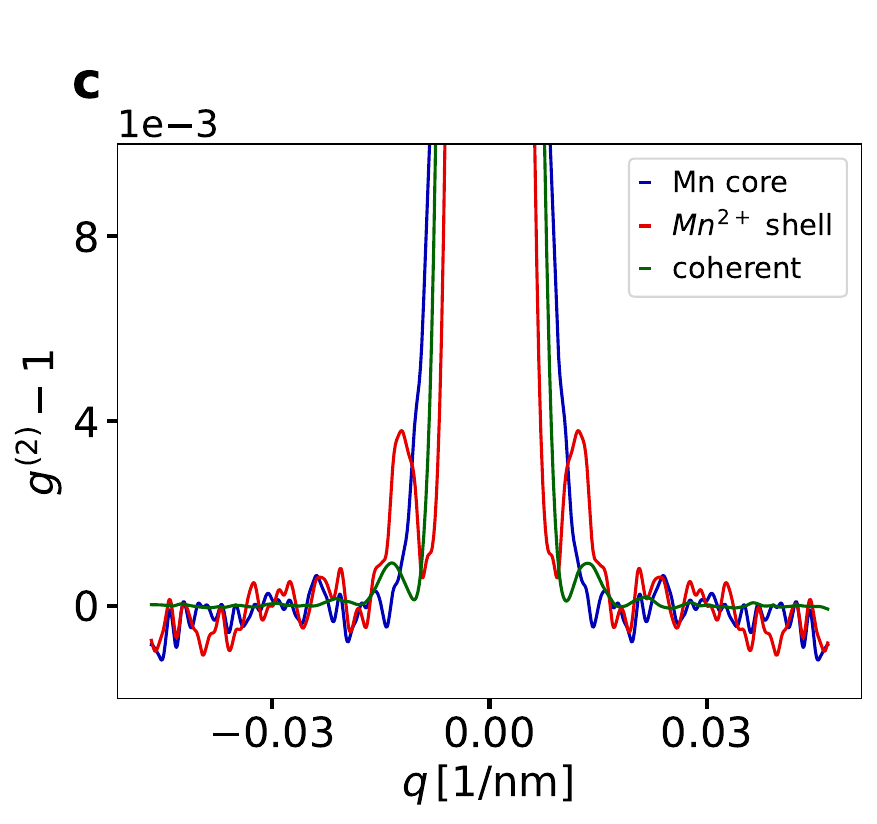} 
  \end{tabular}
\end{tabular}
\caption{K$_\beta$ simulations to study a sample with multiple oxidation states. (\textbf{a}) Integrated 2D detector image showing the broadened K$_\beta$ lines as well as coherent elastic scattering. One can already see the structure of the whole particle reflected in the coherent signal (dark profile to the right), while the fluoresence yields a flat profile (light green). Th (\textbf{b}) Column-wise $g^{(2)}(q, E)$ intensity correlations, normalized by the column-wise standard deviation. Since the spatial distribution of each Mn subspecies differs (see inset in the upper right of (\textbf{a})), one can see different signals within the K$_\beta$ line. (\textbf{c}) The $q$-dependent line profiles for the two oxidation states can be used to reconstruct the structures of each state independently.}
\label{fig:kbeta}
\end{figure}

Let us now consider the case of a heterogeneous nanoparticle comprising a Mn sphere encapsulated in a MnO or similar $\text{Mn}^{2+}$ shell (see sample in Fig.~\ref{fig:kbeta}a). The Si analyzer and the detector are positioned to capture the $\text{K}_{\beta1,3}$ lines as well as the elastic scattering ($E_{\text{elastic}} = \SI{6580}{\eV}$) at a detector distance of $\SI{1.5}{\meter}$. By exploiting the atomic form factor of elastic scattering, its relative intensity can be tuned by placing the Si crystal near-normal to the incident $\mathbf{k}_0$ vector. We estimate 160 photons per exposure due to the lower fluorescence yield of the $\text{K}_{\beta1,3}$ line as well as the reduced solid angle of the detector (see Appendix \ref{app:signal}). We simulated $10^6$ exposures, which would correspond to about 1 hour of data acquisition at the European XFEL~\cite{trost2023imaging}. 

The integrated intensity in Fig.~\ref{fig:kbeta}a shows both the K$_\beta$ lines as well as the coherent diffraction of the entire particle due to its insensitivity to the oxidation state. The spectral shift for the $\text{K}_{\beta1,3}$ line from neutral Mn metal to the $\text{Mn}^2+$ oxidation state is $\SI{1.7}{\eV}$~\cite{peng1994high}. However, since the shift is less than the width of the individual emission lines, the integrated intensity exhibits only one broadened K$_{\beta1,3}$ emission line.

Given that the emission for the two different $\text{K}_{\beta1,3}$ lines arises from two distinguishable materials of distinct shapes, it is reasonable to perform the $g^{(2)}$ calculation solely along the $q$-direction. Additionally, a 2D analysis would be counterproductive here, as it would include correlations from the elastic scattering that contribute only to the offset in the second-order correlation function (see Eq.~\ref{eq:g2}). 

In Fig.~\ref{fig:kbeta}b the one-dimensional $g^{(2)}(q,E)$ profiles for the different energies are shown. Notably, discernible differences along $\mathbf{q}$ are evident for different energies within the broadened K$_{\beta1,3}$ line. The different central peak widths reflect the different shapes of the underlying emitter distributions in the $\text{Mn}-\text{Mn}^{2+}$ compound contributing to each emission line. Higher order features in the $q$ direction contain information about finer structural features. For the elastic scattering, the $g^{(2)}$ profile remains flat and does not provide any additional information, as expected. Figure~\ref{fig:kbeta}c shows the $g^{(2)}$ vs $q$ line profiles. One can clearly see the different Fourier amplitudes corresponding to the core and shell for the two K$_{\beta1,3}$ energies. 

In an experiment where SIDI is combined with forward scattering for coherent diffraction, the  detector can also be positioned farther away from the interaction region to enhance the energy resolution and, particularly, to increase the splitting for the the two K$_{\beta1,3}$ lines. This would improve the $g^{(2)}$-contrast on the one hand, but also decrease the solid angle of the detector leading to a lower photon count per exposure and thus a lower SNR. One can also envision using a different analyzer crystal geometry for higher energy resolution, efficiency or both~\cite{zimmermann2020modern}.

In this letter, we have proposed a new imaging technique capable of performing diffractive imaging of structures with heterogeneous oxidation states for each spectral state independently. Using the high pulse energies and short pulse durations of XFEL sources, IDI~\cite{classen2017incoherent,trost2023imaging} is combined with a spectral analyzer similar to XES or RIXS to perform what we term a spectral incoherent diffractive imaging {SIDI} experiment. Consequently, SIDI provides additional insights beyond the conventional approach of combining coherent diffraction and XES, where the spectrum and the total emitter distributions are acquired, yet the precise emitter configuration inside the scattering particle is not revealed.

This enhanced insight comes at the expense of sacrificing one spatial-dimension in comparison to conventional IDI. Utilizing a single detector restricts imaging to one dimension, but owing to the isotropic nature of fluorescent emission, the detector can be positioned in any suitable orientation and a second and even third crystal-detector pair can be used to probe the orthogonal spatial dimensions. The superior energy resolution of SIDI, in comparison to conventional IDI, not only results in improved $g^{(2)}$-contrast but also expands the maximum feasible transverse sample size to beyond a micrometer, overcoming a seemingly fundamental limit in conventional IDI
 set by the small coherence length and high signal requirements~\cite{lohse2021incoherent}. 

For single particle imaging applications where each pulse exposes a different particle, coherent diffraction can be used to solve for the particle orientation in case of non-spherical samples~\cite{ayyer20213d}. This challenge cannot solely be solved with IDI, as the $g^{(2)}$-contrast is typically too low in a single shot. With each two-dimensional coherent diffraction pattern, additional one-dimensional $g^{(2)}(q,E)$-profiles are obtained, sensitive to the electronic configurations of the sub-domains within a heterogeneous particle. With the promise of \SI{100}{\micro\joule} scale attosecond hard X-ray pulses, the range of applicable systems and spectral sensitivity is only going to increase.

Our proposed method opens avenues for time-resolved, element specific and oxidation state-specific imaging of electron transfer in 3d-transition metal compounds or to study heterogeneous catalysts and battery materials where the nanoscale spatial distribution of elemental oxidation states are crucial for understanding function.  

\begin{acknowledgments}
This work is supported by the Cluster of Excellence 'CUI: Advanced Imaging of Matter' of the Deutsche Forschungsgemeinschaft (DFG) - EXC 2056 - project ID 390715994.
\end{acknowledgments}

\appendix
\section{Simulation details}
\label{app:simulation}
The incident X-ray pulse was simulated as a Gaussian function with a full width at half maximum of $\SI{6.2}{\femto\second}$. Depending on the energy resolution of the actual sample-detector configuration, the pulse was binned into different temporal modes $M$. The photon occupancy in each mode was then calculated from the Gaussian distribution while allowing for $80 \, \%$ variations to account for SASE beam fluctuations.
The sample was approximated as a $\SI{300}{\nano \meter}$ sphere, in the heterogeneous case with an inner core of $\SI{215}{\nano \meter}$ of Mn and a $\SI{85}{\nano \meter}$ shell of $\text{Mn}^{2+}$. 

Based on the signal level (see Appendix~\ref{app:signal}), a random subset of $N_M = N/M$ emitters was chosen for the different temporal modes. At the detector, each atom contributes a plane wave with a random initial phase $\phi_{j,m} \in [0, 2\pi)$:

\begin{equation}
    \psi_m(\mathbf{k}) \propto \sum_{j=1}^{N_M} e^{i (\mathbf{k} \mathbf{r}_{j,m} + \phi_{j,m})},
\end{equation}
where $\mathbf{r}_{j,m}$ is the position of the emitter $j$ in mode $m$ and $\mathbf{k}$ the corresponding scattered wave vector $\mathbf{k} = \mathbf{k_\text{out}} - \mathbf{k_0}$. Thus, each detector pixel can be assigned to a specific $\mathbf{k}$. The length $|\mathbf{k}|$ was assumed to be constant as the variation was less than 1 pixel across the energy range of the detector.

In order to account for the angular dispersion of the Si crystal for different photons energies, the resultant single-pixel-wide intensity line on the detector was convolved by the normalized Lorentzian spectral profile $\tilde S(E)$ of the emission lines:

\begin{equation}
    I(k,E) = \sum_{m=1}^{M} \tilde S(E) |\psi_m(k) |^2 
\end{equation}
where the sum is over each temporal mode.

Finally, the intensity was Poisson-sampled to simulate single photon detection. Additional photons from background scattering were neglected as they would distribute uniformly across the detector while the actual signal is concentrated along the emission profiles. Therefore, even with background only a slight reduction of contrast and SNR is expected.

The reconstruction was then implemented by calculating the sum of the intensity autocorrelations for each individual frame, normalized by the autocorrelation of the integrated intensity over all frames. The code for both simulation and $g^{(2)}$ calculation are available at \texttt{https://github.com/TammeWollweber/spectralIDI}.

\section{Signal estimation}
\label{app:signal}
The intensity $I(k,E)$ on the detector for a particular fluorescence channel in number of photons can be calculated as follows:

\begin{equation}
    I(k, E) = I_0 \cdot \sigma_\text{abs} \cdot \Phi \cdot N \cdot \Omega \cdot \tilde{S}(E),
    \label{eq:signal}
\end{equation}

where $I_0$ is the incident fluence, $\sigma_\text{abs}$ is the absorption cross-section for a single atom, $\Phi$ is the fluorescent yield of the given emission line, $N$ is the number of scattering atoms and $\Omega$ is the effective solid angle of the detector, taking into account that the silicon crystal is placed $\SI{300}{\milli\meter}$ from the interaction point and that only a small strip of the crystal is reflective for a given energy. $\tilde{S}(E)$ is again the normalized Lorentzian spectral profile. The intensity is independent of the vertical position on the detector, $k$.

We considered a $\SI{100}{\micro\Joule}$ X-ray pulse focussed to a \SI{300}{\nano\meter} spot, yielding a fluence of \SI{1.34e12}{ph/\micro\meter\squared}. This yields on average 5.47 absorption events per atom~\cite{henke1993x}, so we assumed saturated absorption i.e. $I_0 \cdot \sigma_\text{abs} = 1$ in Eq.~\ref{eq:signal}.

For a $\SI{300}{\nano \meter}$ particle, the number of absorbing atoms is \SI{2.6e9}{}. The tabulated fluorescence yield values of $\Phi_\alpha = 0.27$ and $\Phi_\beta = 0.025$ were used for each simulation~\cite{elam2002new}. The effective solid angles of the detector are $\Omega_\alpha = \SI{5.76e-6}{\sterradian}$ and $\Omega_\beta = \SI{2.56e-6}{\sterradian}$ for a detector distance of \SI{1}{\meter} and \SI{1.5}{\meter}, respectively.

\bibliography{lib}

\end{document}